\documentstyle[aps,prl,preprint]{revtex}
\draft
\begin{document}
\title{ Generating Converging Bounds to the (Complex) Discrete States of
the $P^2 + iX^3 + i\alpha X$ Hamiltonian}
\author{C. R. Handy}
\address{Department of Physics \& Center for Theoretical Studies of 
Physical Systems, Clark Atlanta University, 
Atlanta, Georgia 30314}
\date{Received \today}
\maketitle
\begin{abstract}
The Eigenvalue Moment Method (EMM), Handy (2001), Handy and Wang (2001))
 is applied to the $H_\alpha \equiv P^2 + iX^3 + i\alpha X$
Hamiltonian, enabling the algebraic/numerical generation of
 converging bounds to the complex energies of the
$L^2$ states,
 as argued (through asymptotic methods) 
by Delabaere and Trinh (J. Phys. A: Math. Gen. {\bf 33} 8771 (2000)). 
The robustness of the formalism, and its computational implementation, 
suggest that the present nonnegativity formulation  implicitly contains
the key algebraic relations by which to prove Bessis'
conjecture that the eigenenergies of the
$H_0$ Hamiltonian are real. 
 The required algebraic analysis of the
EMM procedure pertaining to this problem will be presented in 
a forthcoming work.
\end{abstract}
\vfil\break
\section{Introduction}
\subsection {General Overview}
There has been much speculation on the mechanism responsible for 
symmetry breaking within a special class of ${\cal P}{\cal T}$ invariant
Hamiltonians. It has been argued by Bender and Boettcher (1998), based on
a conjecture by D. Bessis, that
the class of Hamiltonians of the form $P^2 + (iX)^n$ admits bound states,
within the complex plane, with real discrete spectra. Their arguments
show that the ${\cal P}{\cal T}$ invariance of the Hamiltonian
 is reflected in the
wavefunction, $\Psi^*(-x) = \Psi(x)$, resulting in real spectra.
However, Delabaere and Trinh (2000) have emphasized that ${\cal P}{\cal T}$
invariance of the Hamiltonian is not sufficient to prevent symmetry 
breaking solutions. For instance,
the Hamiltonian $P^2 + iX^3 + i\alpha X$ admits
bounded ($L^2$) solutions on the real axis, which can have complex 
energies for $\alpha < \alpha_{critical} < 0$
 (thereby breaking ${\cal P}{\cal T}$ 
invariance), or real energies, for $\alpha > \alpha_{critical}$.

An understanding of the underlying mechanism for symmetry breaking has remained
elusive, despite the numerous investigations on the above, and 
related problems, by Bender, Boettcher, and Meisinger (1999),
Bender et al (1999), Bender et al (2000), Bender and Wang (2001), Caliceti (2000), Delabaere and Pham (1998), Handy (2001), Handy and Wang (2001), 
Levai and Znojil (2000), Mezincescu (2000,2001),
Shin (2000), and Znojil (2000), in addition to those already cited.
 However, the recent work by  Dorey, Dunning, and Tateo (2001) presents
one possible explanation.

Our objective is to seek alternative (and less analytical) arguments that can possibly shed some light on this matter. In this regard, we have attempted to implement a novel ``positivity quantization" formalism 
based on the recent works by Handy (2001) and Handy and Wang (2001). These 
in turn make use of the Eigenvalue Moment Method (EMM) originally developed by
Handy and Bessis (1985) and Handy et al (1988a,b).

Our results, as communicated here, are very impressive. We are able to generate converging 
bounds for the (complex) discrete state energies, and arbitrary $\alpha$. 
We are able to confirm the general results derived by Delabaere and Trinh, 
although our methods do not rely on asymptotic estimates, as theirs do.

It is important to emphasize that the EMM approach generates an infinite hierarchy of (closed) algebraic inequalities. We investigate the consequences of these relations from a numerical perspective. This is an important first step in 
identifying the algebraic relations responsible for symmetry breaking solutions, within our formalism. Our results strongly suggest that a careful algebraic analysis of the underlying EMM relations will serve to identify the theoretical structure leading to a
proof of Bessis' conjecture. This will be presented in a future communication.
We emphasize that unlike numerical integration schemes, which do not necessarily provide an undestanding of the underlying theoretical structure of a problem,
the algebraic/numerical structure of EMM can.

Beyond this, the ability to generate converging bounds to complex energy levels is a remarkable feat in its own right. This motivates the present communication. In a related work by Handy, Khan, Wang, and Tymczak (HKWT, 2001),
 they show how the
{\it Multiscale Reference Function} formulation (Tymczak et al (1998a,b))
 can easily generate the (complex) discrete state energies, for arbitrary $\alpha$. Their estimation methods 
generate energies that fall within the bounds given here. As such, the present 
bounding theory provides a confidence test for the, numerically faster, 
MRF method. 
 Because of this, we only generate energy 
bounds for important $\alpha$ values near the first complex-real bifurcation
point, $\alpha_{critical} = -2.6118094$, as predicted by MRF.
\vfil\break

\subsection{Technical Overview}

The 
recent work by  Handy (2001) introduced a new formalism that transforms the
one dimensional Schrodinger equation into a fourth order, linear differential equation for $S(x) \equiv |\Psi(x)|^2$, regardless of the (complex) nature of
the potential, $V(x) = V_R(x) + i V_I(x)$:

\begin{eqnarray}
-{1\over {V_I-E_I}}S^{(4)} - \Big({1\over{V_I-E_I}}\Big)'S^{(3)}
+ 4 \Big( {{V_R-E_R}\over {V_I-E_I}} \Big) S^{(2)} \cr
 + \Big( 4\Big({{ V_R-E_R}\over { V_I-E_I}}\Big)'+ 2\Big({{{V_R}'}\over
{ V_I-E_I}}\Big )
          \Big) S^{(1)}
+ \Big( 4  (V_I-E_I) + 2\Big({{{V_R}'}\over { V_I-E_I}}\Big)'\Big) S = 0,
\end{eqnarray}
where $S^{(n)}(x) \equiv \partial_x^nS(x)$, and $E = E_R+iE_I$, etc.
Through this fourth order equation, one is able to transform
the quantization problem into a nonnegativity 
representation suitable for a Moment Problem (Shohat and Tamarkin (1963)) based analysis, utilizing the Hamburger moments
\begin{equation}
u_p = \int_{-\infty}^{+\infty} dx \ x^p S(x).
\end{equation}

Such methods were originally developed by Handy and Bessis (1985), and Handy et al (1988a,b), and used to generate rapidly converging bounds to the bosonic 
ground state energy of singular perturbation/strong coupling (multidimensional)
systems. This moment problem based, ``positivity quantization", approach is
generically referred to as the Eigenvalue Moment Method. An 
efficient algorithmic
implementation requires the use of linear programming (Chvatal (1983)). 

We emphasize that the only bounded (i.e. $L^2$ functions within the $\Psi$-representation, or $L^1$, within the $S$-representation) and 
nonnegative solutions to Eq.(1) 
are the physical solutions (Handy (2001)). This is because if 
$\Psi_{1;E}(x)$ and $\Psi_{2;E}(x)$ are independent solutions 
of the Schrodinger equation, for arbitrary $E$
 (and thus unbounded, except for the
physical solutions),  then 
$|\Psi_{1;E}(x)|^2$, $|\Psi_{2;E}(x)|^2$,
$\Psi_{1;E}^*(x) \Psi_{2;E}(x)$, and $\Psi_{1;E}(x) \Psi_{2;E}^*(x)$,
are the independent solutions to Eq.(1), assuming $V$ is complex. If
$V$ is real, then all the solutions  are real, and only
the first three configurations are independent; leading to 
a third order linear differential equation for $S$ (Handy (1987a,b)).  Thus,
because of the uniqueness of nonnegative and bounded solutions,
application of  EMM to the relevant moment 
equations (i.e. Eqs.(47 \& 51) for Eq.(1),
 or the alternate moment formulation discussed
in Sec. II) will generate converging bounds to the physical energies 
(Handy and Bessis (1985), Handy et al (1988a,b)).

Despite this, we discover that
for problems with complex energies, $E_I \neq 0$, the derivation of,
as well as the actual,
 moment equation obtained
from Eq.(1), are  not the most efficient.
 A more efficient generation of the required $u$-moment equation  can be derived by working within a broader framework involving three coupled differential equations for the probability density, $S(x)$, the kinetic energy density function, $P(x) \equiv |\Psi'(x)|^2$, and the probability current density, $J(x) = 
{{\Psi(x)\Psi'^*(x) - \Psi^*(x)\Psi'(x)}\over {2i}}$. The first two expressions
are nonnegative configurations. In particular cases, the probability current 
density will also be nonnegative. 
 This is discussed in Sec. II.
However, this is not the immediate focus
of the present work. Instead, it is to show how the realization of
a nonnegative, linear, differential representation for $S$ leads to a very 
effective quantization procedure, capable of confirming the existence 
of symmetry breaking solutions.

Working within the coupled system of equations for $\{S,P,J\}$,
 one can generate many more moment constraints for the $E_I \neq 0$ case. It is in this sense that we say that
the moment equation derived directly from Eq.(1) 
is ``incomplete", when $E_I \neq 0$. Generating as many moment constraints as possible speeds up the convergence rate of the bounds. However, as
noted above, even if we work with the reduced set  of moment constraints generated from Eq.(1), the bounds generated will converge to the unique physical answer, just more slowly.

Working directly with Eq.(1) leads to a complete set of moment equation constraints when $E_I = 0$.
This is the case examined by Handy (2001), corresponding to the
$P^2 + (iX)^3$ problem (assuming real
spectra, Handy (2001)).
Application of EMM analysis in this
case yielded impressive bounds for the first five energy levels (Handy (2001)),
as well as complex rotated versions of the Hamiltonian (Handy and Wang (2001)).

The sensitivity of the EMM procedure, as evidenced by the excellent nature of
these bounds, strongly suggest that Bessis' conjecture is more likely a 
consequence of some underlying algebraic identity, than a more subtle analytic
constraint.

Application of EMM to the more general case, $E_I \neq 0$, 
also yields  very good
bounds to the complex eigenenergies, as shown in this work. We have implemented
our bounding analysis up to a relatively low moment order. Our objective has 
been to affirm the $S-$EMM formalism's relevancy as an effective bounding
theory.

\vfil\break
\section{Derivation of the $S$-Moment Equation}

We  derive Eq.(1) in a manner different from that  presented by Handy (2001). As noted previously, the present formalism is better suited for problems
with $E_I \neq 0$.

Denote the  Schrodinger equation by

\begin{equation}
H_x \Psi(x) = E\Psi(x),
\end{equation}
where the normalized Hamiltonian is  $H_x = -\epsilon\partial_x^2 + V(x)$, 
involving a complex potential, $V$, and complex energy, $E$. We make explicit the kinetic energy ``expansion" parameter,
$\epsilon$. As recognized by Handy (1981), within a moments representation,
kinetic energy expansions become analytic.
 This will be an important component of
the present theory.
The following discussion assumes that $\epsilon$ is real and positive.

It is readily apparent that if $V^*(-x) = V(x)$, and $\Psi(x)$ is a bound state with complex energy, $E$, then $\Psi^*(-x)$ is another bound state solution with energy $E^*$.
Thus,  complex roots  come in complex conjugate pairs, for ${\cal P}{\cal T}$
invariant Hamiltonians.

\subsection {The Coupled  Differential Equations for $ |\Psi(x)|^2$, $|\Psi'(x)|^2$, and ${{\Psi(x)\Psi'^* (x) - \Psi^*(x)\Psi'(x)}\over {2i}}$ }
Define the quantities $S(x) = |\Psi(x)|^2$, $P(x) = |\Psi'(x)|^2$, and
$J(x) = {{\Psi(x)\Psi'^* (x) - \Psi^*(x)\Psi'(x)}\over {2i}}$. 
The expressions $\Sigma_1(x) \equiv \Psi^* H_x\Psi(x) + c.c.$,
$\Delta_1(x)  \equiv \Psi^* H_x\Psi(x) - c.c.$, and 
$\Sigma_2(x)  \equiv \Psi'^* H_x\Psi(x) + c.c.$, satisfy (Handy and Wang (2001))

\begin{equation}
\Sigma_1(x) =  \epsilon(2P(x) - S''(x)) + 2 (V_R(x) - E_R)S(x) = 0,
\end{equation}
\begin{equation}
\Sigma_2(x) = -\epsilon P'(x) + (V_R(x) - E_R)S'(x) -2 (V_I(x) - E_I) J(x) = 0,
\end{equation}
and
\begin{equation}
-{i\over 2}\Delta_1(x) = (V_I(x) - E_I)S(x) + \epsilon\partial_xJ(x) = 0.
\end{equation}

For the potential in question, $V(x) = ix^3+i\alpha x$, if $S(x),P(x),J(x)$
form a solution set corresponding to energy $(E_R,E_I)$, then 
$S(-x),P(-x),J(-x)$ is a solution set for energy $(E_R,-E_I)$.

In obtaining Eq.(1) (i.e. $\epsilon = 1$), we differentiate Eq.(4)
 and use Eq.(5)
to substitute for $P'$. Upon dividing the resulting expression by $V_I-E_I$,
and differentiating, we use Eq.(6) to substitute for $J'$. For 
future reference, we make this process explicit. 

All of the physical configurations $\{S,P,J\}$ are implicitly 
assumed to be bounded and vanish at infinity. Thus,

\begin{equation}
J(x) = -{(\epsilon\partial_x)}^{-1}\Big((V_I(x) - E_I)S(x)\Big ),
\end{equation}
where $\partial_x^{-1} \equiv \int_{-\infty}^x dx_v$. In addition,
\begin{equation}
P[S;x] = {1\over{2}} S''(x) -  {1\over \epsilon}(V_R(x) - E_R)S(x).
\end{equation} 
We then obtain:
\begin{equation}
 -\epsilon^2 \partial_xP[S(x);x] + \epsilon 
(V_R(x) - E_R)S'(x) +2 (V_I(x) - E_I) {(\partial_x)}^{-1}\Big((V_I(x) - E_I)S(x)\Big ) = 0.
\end{equation}
In order to transform this into Eq.(1), we simply apply the operator
$\partial_x {{1\over{V_I(x)-E_I}}}$, resulting in a fourth order, linear,
homogoneous, diffential equation.

\subsection{The $S$-Moment Equation}
Let us denote by $\{u_p,v_p,w_p\}$ the Hamburger moments of the three functions 
$S,P,J$, respectively.  Thus, $u_p \equiv \int_{-\infty}^{+\infty} dx \ x^p S(x)$, etc.,
for $p \geq 0$. For the present problem, $V_R = 0$, and $V_I = x^3+\alpha x$.
Multiplying each of the three equations by $x^p$, and integrating by parts,
yields

\begin{equation}
2\epsilon v_p - p(p-1)\epsilon u_{p-2} - 2 E_R u_p = 0,
\end{equation}
\begin{equation}
p\epsilon v_{p-1} + E_R p u_{p-1} - 2 (w_{p+3} + \alpha w_{p+1} - E_Iw_{p}) = 0,
\end{equation}
and
\begin{equation}
(u_{p+3} + \alpha u_{p+1} - E_Iu_p) - p\epsilon w_{p-1} = 0,
\end{equation}
$p \geq 0$. Note that from Eq.(10) (for $p = 0$)  $E_R$ must
be positive.

We can convert this into a moment equation for the $\{u_p\}$'s by 
using the first moment relation to solve for $v_p$ in terms of the $u$'s.
Likewise, taking $p \rightarrow p+1$ in the last moment relation, determines
$w_p$ in terms of the $u$'s. Finally, substituting both relations in 
the second equation generates a moment equation for the $u$'s:

\begin{eqnarray}
\epsilon^2 (p - {{3p^2}\over 2} + {{p^3}\over 2}) u_{p-3}
+2p\epsilon E_R u(p-1) - {{2E_I^2}\over {(p+1)}} u_{p+1} 
+{{2\alpha E_I}}\Big( {1\over{p+1}} + {1\over{p+2}}\Big) u_{p+2}
-{{2\alpha^2}\over{p+2}} u_{p+3} \cr
+2E_I \Big( {1\over{p+1}} + {1\over{p+4}}\Big) u_{p+4}
-2\alpha\Big({1\over{p+2}} + {1\over{p+4}}\Big) u_{p+5}
-{2\over{p+4}}u_{p+7} = 0.
\end{eqnarray}
This moment equation holds for all $\epsilon$, including $\epsilon = 0$.
In the latter case, upon multiplying the $u$-moment equation by 
$(p+1)(p+2)(p+4)$, the resulting relation (quadratic in $p$) 
incorporates the relation given in Eq.(12) for $\epsilon = 0$.

We note that  the $u$-moment equation does not
include one important additional moment constraint, that for the
$p = 0$ relation in Eq.(12). This yields

\begin{equation}
u_{3} + \alpha u_{1} = E_Iu_0  .
\end{equation}
Of course, this is the relation one obtains directly from the
Schrodinger equation, upon multiplying it by $\Psi^*(x)$:
\begin{equation}
\int dx \ (\epsilon P(x) + V(x) S(x)) = E \int dx \ S(x),
\end{equation}
and identifying the real and imaginary parts (the latter corresponding
to Eq.(14)).

The recursive, linear, homogeneous,
structure of the $u$-moment equation tells us that
all of the moments are linearly dependent on the first seven moments
$\{u_\ell|0 \leq \ell \leq 6\}$. The additional constraint in Eq.(14)
allows us to solve for $u_3$ in terms of $u_0$ and $u_1$.
Thus, the reduced set of independent moments is  $\{u_\ell| 0 \leq \ell \leq 2,
4 \leq \ell \leq 6\}$. These are referred to as the {\it missing moments}.
In addition, we must impose a suitable normalization condition. The details
are given in the next section.

There is an important theoretical point that must be stressed. The
recursion relation in
Eq.(13) is the moment equation resulting from integrating both
sides of Eq.(9) by $x^p$. Both Eq.(9) and Eq.(1) are 
equivalent to each other, as explained above. However, the
moment equation generated by Eq.(1) (refer to Eq.(47) and Eq.(51))
is different from 
Eq.(13), as explained in the Appendix. It will involve one 
more degree of freedom (i.e. {\it missing moment} order)
than Eq.(13). However, since Eq.(1) and Eq.(9) are 
equivalent, and Handy (2001) has argued that the only bounded
and nonnegative solutions are those corresponding to the 
physical configurations, application of EMM to either will 
result in converging bounds. Naturally, the moment equation 
involving fewer independent variables (i.e. {\it missing moments})
will yield faster converging bounds. 

In addition, because Eq.(13) is the moment equation of Eq.(9),
and similarly for Eqs.(47 \& 51) and Eq.(1),
any additional constraints, such as Eq.(14),
are not required in order to generate converging bounds. Such 
constraints only improve the convergence rate of the bounds by
reducing the number of independent, missing moment, variables.

>From a different perspective,
the manifest difference between Eq.(13) and  Eq.(47 \& 51)
hinges on the fact that in obtaining Eq.(1)
we implicitly take $p \rightarrow p+1$ in 
Eq.(11). We note that this implies that Eq.(1) cannot generate   the
extra constraint $w_3+\alpha w_1 - E_I w_0 = 0$, that is,
the $p = 0$ relation from Eq.(13).

If $E_I = 0$, and the nonnegative
configurations $S(x)$ and $P(x)$ are symmetric, then
these extra constraints (including Eq.(14)) are nonexistent. That is,
Eq.(1) yields a complete set of moment constraints if $E_I = 0$.

The moment equations in Eqs.(13-14) are the preferred relations ( because
they are easier to derive, and involve less missing moments),  if $E_I \neq 0$.

We will work with the $u$-moment equation as given above, complemented
by Eq.(14).

\subsection {The Zeroth Order $\epsilon$-Contribution}

As previously noted, one of the most important reasons for working within a moments' representation is
that it is analytic in $\epsilon$.

If $\epsilon \neq 0$, there is equivalency between the
 sets of equations  Eq.(10-12) and Eq.(13-14). When $\epsilon = 0$
 (and thus $E_R = 0$), the 
$w$ moments decouple from the $u$ moments. We find that  Eq.(13-14) 
includes more solutions than those generated from Eq.(12). 
However, not all of these will be consistent with EMM quantization.
 One formal way
of understanding this is to consider Eq.(9) when $\epsilon = 0$. We obtain
\begin{equation}
 (V_I(x) - E_I) {(\partial_x)}^{-1}\Big((V_I(x) - E_I)S^{(0)}(x)
\Big ) = 0.
\end{equation}

For real, bounded, configurations, $\{f(x),g(x)\}$, the integral 
$\int_{-\infty}^{+\infty} dx \ f(x) (\partial_x)^{-1} g(x)$ becomes 
$\int_{-\infty}^{+\infty} dx\ \Big(((\partial_x)^{-1})^\dagger f(x)\Big ) \ g(x)$, where
$((\partial_x)^{-1})^\dagger \equiv \int_x^{+\infty}dx_v$. This follows from
$\int_{-\infty}^{+\infty}dx\ f(x) \int_{-\infty}^xdx_v \ g(x_v) = 
\int_{-\infty}^{+\infty}dx\ \int_{-\infty}^0d\xi\ f(x) \ g(\xi+x) = 
\int_{-\infty}^{+\infty}dx\ \int_{-\infty}^0d\xi\ g(x) f(x-\xi) = 
\int_{-\infty}^{+\infty}dx\ g(x) \int_{x}^{+\infty} dx_v\ f(x_v)$.

Multiply Eq.(16) by $x^p R_\beta(x)$, where $R_\beta(x)$ is a regulating (bounded) function which reduces to unity when $\beta \rightarrow 0$. Integrating, and using   $\Big( (\partial_x)^{-1}\Big )^\dagger$, gives (i.e. in the $\beta \rightarrow 0$ limit)
\begin{equation}
((\partial_x)^{-1})^\dagger\Big(x^{p+3} + \alpha x^{p+1} - E_I x^p \Big)  \rightarrow
{1\over{p+4}}x^{p+4} + {\alpha\over{p+2}} x^{p+2} - {{E_I}\over {p+1}}x^{p+1}.
\end{equation} 
This, in turn, upon multiplying by $(V_I(x) - E_I)S^{(0)}(x)$, and
completing the $x$-integration, gives the zeroth order (in $\epsilon$) moment
relation in Eq.(13).

Therefore, the $\epsilon = 0$ moment equation in Eq.(13) (i.e.
 corresponding to Eq.(16)) 
tells us that

\begin{equation}
(V_I(x) - E_I) J^{(0)}(x) = 0.
\end{equation}
This has the general, formal, solution $J^{(0)}(x) = \sum_\ell {\cal J}_\ell
\delta(x-\tau_\ell)$, where the $\tau_\ell$'s are the turning points:
\begin{equation}
V_I(\tau_\ell(E)) = E_I.
\end{equation}
The ${\cal J}_\ell$'s are arbitrary. However, the $\epsilon = 0$ solution
to Eq.(12) really correspond to $J = 0$. We can (formally) argue this
by applying
 $\partial_x (V_I(x)-E_I)^{-1}$ to Eq.(18), yielding
\begin{equation}
\partial_xJ^{(0)}(x) = (V_I(x) - E_I) S^{(0)}(x) = 0,
\end{equation}
which is the underlying configuration space relation corresponding to
Eq.(12), for $\epsilon = 0$.
The only possible, bounded, solution is $J = 0$.

Thus, the only solution to Eq.(13), for $\epsilon = 0$, consistent
with the EMM quantization constraints (which demand boundedness and
nonnegativity), as discussed in the following section, should be
that corresponding to $J^{(0)}(x) = 0$.

Consistent with the previous discussion,
the $\epsilon = 0$ moment equations (Eq.(10-12)) yield
$E_R = 0$,
\begin{equation}
w_{p+3}^{(0)} + \alpha w_{p+1}^{(0)} - E_Iw_{p}^{(0)} = 0,
\end{equation}
and
\begin{equation}
u_{p+3}^{(0)} + \alpha u_{p+1}^{(0)} - E_Iu_{p}^{(0)} = 0,
\end{equation}
$p \geq 0$.
The solution set to these are $u_p = \sum_\ell {\cal A}_\ell \tau_\ell^p$,
and $w_p = \sum_\ell {\cal J}_\ell \tau_\ell^p$. 

All of the (complex) turning points, for
arbitrary $E_I$, contribute to the zeroth order structure of $S(x)$.

\subsection {The Moment Problem Constraints}

The Moment Problem
conditions for nonnegativity (Shohat and Tamarkin (1963)) correspond to the
inequalities:

\begin{equation}
\int dx \ (\sum_{j=0}^NC_n x^n)^2 S(x) \geq 0,
\end{equation}
for $N < \infty$ and $C_n$ arbitrary, real,  variables.
 These are usually transformed into  nonlinear (in the moments)
Hankel-Hadamard (HH) determinantal inequalities, $\Delta_{0,N}(u) > 0$,
where $\Delta_{0,N}(u) \equiv Det(u_{i+j})$, and $0 \leq i,j \leq N$.
 However, we prefer
their linear (in the moments)
 equivalent, corresponding to the quadratic form inequalities:
\begin{equation}
\sum_{n_1,n_2 = 0}^N C_{n_1} u_{n_1+n_2} C_{n_2} \geq 0.
\end{equation}
These will be referred to as the linear HH nonnegativity
constraints.

The moment (``Hankel") matrix $u_{n_1+n_2}$ is linearly
 dependent
on the {\it missing moments}, $\{u_0,u_1,u_2,u_4,u_5,u_6\}$,  
 nonlinearly dependent on 
 $(E_R,E_I)$, and analytic with respect to $\epsilon$.
With respect to the following discussion, we note that
 through an appropriate normalization prescription (as discussed in 
Sec. III), the missing moment variables will lie within a bounded, convex,
domain.

The physical solution must satisfy all of the above constraints. To
any finite order, $N$, if at a given energy value, $E = (E_R,E_I)$, 
there is a missing moment solution set, ${\cal U}_{N;E}$,
 then it must be convex.
The EMM eigenenergy bounding procedure simply involves determining the
energy subregions, $(E_R,E_I) \in {\cal R}_{N;j}$, for which ${\cal U}_{N;E}$
exists. The $j$-index  enumerates the discrete states. 

Based on the many applications of EMM over the last sixteen years, one
expects the ${\cal R}_{N;j}$ regions to be connected and bounded, although not
necessarily convex. This is supported
 by our empirical results.

The boundary of the smallest rectangle containing ${\cal R}_{N;j}$
(i.e. $[E_R^{(L)},E_R^{(U)}] \times [E_I^{(L)},E_I^{(U)}] 
\supset {\cal R}_{N;j}$) define
the bounds for $(E_R,E_I) \in {\cal R}_{N;j}$. In practice, we numerically 
determine a slightly larger rectangle than this.
Our numerical analysis will be based on the above relations. The
following discussion addresses an interesting side issue that enhances
our understanding of the above relations. 

The numerically minded reader may wish to skip to Sec. III.

\subsection {Simplification of the EMM Constraints for Nonnegative $J(x)$}

The Hankel-Hadamard inequalities are
automatically
satisfied by the atomic distribution $S(x) \rightarrow S^{(0)}(x) = \sum_\ell {\cal A}_\ell
\delta(x-\tau_\ell)$, provided ${\cal A}_\ell \geq  0$. The zero equality
is satisfied by polynomials, $P_{N;C}(x) \equiv \sum_{n=0}^NC_nx^n $
  whose roots include all of the turning points,
$P_{N;C}(x) = P_{N-3;{\tilde C}}(x) \times (V_I(x) - E_I)$.

For physical (non-atomic distribution) solutions, only the strict inequality
can be satisfied. In this case, we can work with the relations

\begin{equation}
\int dx \ (\sum_{j=0}^{N}{\tilde C}_n x^n)^2 (V_I(x)-E_I)^2 S(x) > 0.
\end{equation}
Normally, one would prefer to work with such positive relations because they
will not contain any zeroth order $\epsilon$ dependence; thereby 
generating the  positivity constraints that really contribute  to 
quantization. However, in the present case, such inequalities are not
independent of zeroth order $\epsilon$ contributions. 

Instead, as argued before, the zeroth order structure of the moment equation
in Eq.(13) is due to the probability current, $J$. However, we can only work
with nonnegativity constraints for the current, if it is nonnegative.

If the probability current is nonnegative, then
\begin{equation}
\int dx \ (\sum_{j=0}^{N}C_n' x^n)^2 (V_I(x)-E_I)^2 J(x) \geq 0,
\end{equation}
has no zeroth order $\epsilon$ dependence.  That is, from Eq.(18), all the
zeroth order terms are elliminated. This set of constraints is
algebraically simpler than working with the analogue of Eq.(23), 
as applied to $J$, assuming $J \geq 0$.

\subsection{{Properties of the $S,P,J$ Equations and (Minimal) Conditions for $J$'s Positivity }}

We assume that $\epsilon > 0$, $E_R > 0$, and $V_R = 0$.
The turning points satisfy $V_I(\tau_\ell) = E_I$. The following
analysis is restricted to the real axis. We also assume that
all of the configurations have analytic extensions into the
complex-$x$ axis. Thus, $S(x) = \Psi(x) \Psi^*(x^*)$ is the
analytic extension of $|\Psi(x)|^2$, etc. Since $V(x)$ is 
a regular function, the wavefunctions $\Psi(x)$ and $\Psi^*(x^*)$
are regular functions in the complex $x$-plane; hence, so too
are $s(x), P(x),$ and $J(x)$.

Both $S$ and $P$ are nonnegative. Because of their definitions,
both $S$ and $P$ cannot be zero simultaneously, except at
infinity (i.e. otherwise
 $\Psi$ and $\Psi'$
would be zero simultaneously, generating the trivial zero solution).
When $S(x_o) = 0$, or $P(x_o) = 0$, then $\Psi(x_o) = 0$, or
$\Psi'(x_o) = 0$, respectively; therefore,
the probability current, $J$, is zero at all of the zeroes of $S$ and
$P$.

It follows from the nonnegativity of $S$ that if $S(x_o) = 0$, then
 these are extremal points, $S'(x_o) = 0$. The same holds for
 $P$.

\bigskip\noindent {\bf {Lemma \# 1: If $S(x_o) = 0$, 
and $x_o \neq \tau$ ( a real turning point), then $J$ becomes 
negative in the neighborhood of $x_o$ }}

Since $S(x_o) = 0$, then $S'(x_o) = 0$, and $J(x_o) = 0$. 
>From Eq.(6), $J'(x_o) = 0$. Differentiating
Eq.(6) once, we obtain $J''(x_o) = 0$. Differentiating it a second
time yields $-\epsilon J'''(x_o) = (V_I(x_o)-E_I)S''(x_o)$. 
>From Eq.(4) $ 2 P(x_o) = S''(x_o) \neq 0$. Thus, since $x_o \neq \tau$,
it follows that $J'''(x_o) \neq 0$. That is, the local power
series expansion for $J$ becomes $J(x) = {1\over 6}J'''(x_o)(x-x_o)^3
+O((x-x_o)^4).\ \diamond$

\bigskip\noindent {\bf {Lemma \# 2: If $J'(x_o) = 0$,
and $x_o \neq \tau$, then $J$ bececomes
negative in the neighborhood of $x_o$ }}

>From Eq.(6) it follows that $S(x_o) = 0$ and Lemma \#1 applies.$ \ \diamond$

\bigskip\noindent {\bf {Lemma \# 3: If $(V_I(x)-E_I)$ is 
asymptotically monotonically increasing, with one real turning point,
 then $J(x) > 0$. $J(x)$ is strictly increasing for $x < \tau$,
and strictly decreasing for $x > \tau$. }}

By assumption, there is only one real turning point, $\tau$.
Also, $\lim_{x\rightarrow \pm \infty} (V_I(x)-E_I) = \pm \infty$.
>From Eq.(5) we have
  $(\epsilon P + E_R S)' = -2(V_I(x)-E_I)J(x)$; however,
since $S$ and $P$ are asymptotically positive and decreasing to
zero, we have $\lim_{x\rightarrow \pm \infty}(\epsilon P + E_R S)' =
 0^{\mp}$. Accordingly, $\lim_{x\rightarrow \pm \infty}J(x) = 0^+$.
It then follows that  there must be two points (possibly the same),
coming in from $-\infty$ and $+\infty$, where $J$ is a positive local
maximum. Denote these by $x_{o_1} \leq x_{o_2}$. At these points
we have $J'(x_{o_{1,2}}) = 0$. If either of these points is not
$\tau$, then by Lemma \#2, we have a contradiction. Thus 
$x_{o_{1,2}} = \tau$, and $J(x)$ is strictly positive. Clearly,
$J'(x) > 0$, for $x < \tau$, and $J'(x) < 0$, for $x > \tau.\ \diamond$

\bigskip\noindent {\bf {Lemma \# 4: If 
 $\alpha^3 + {{27}\over 4}E_I^2 > 0$, then
$J(x) > 0$}}

Assume $V_I(x) = x^3+\alpha x$. If $\alpha \geq 0$, then $V_I(x)-E_I$ satisfies the conditions of
Lemma \#3. If $\alpha < 0$, then $V_I(x)$ will have  a local
maximum at $x_- = - ({{|\alpha|}\over 3})^{1\over 2}$, and 
a local minimum at $x_+ = -x_-$. At these locations, we have
$V_I(x_-) = {2\over {3^{3\over 2}}}{|\alpha|^{3\over 2}}$,
and $V_I(x_+) = -V_I(x_-)$. If $E_I > V_I(x_-)$, or $E_I < V_I(x_+)$,
then the conditions of Lemma \# 3 are satisfied.$\ \diamond$

\bigskip\noindent {\bf {Lemma \# 5: If $J(x) \geq 0$, then 
its local extrema must occur at the turning points}}

The local extrema correspond to $J'(x_o) = 0$. If $x_o \neq \tau$
(where there can be more than one real turning point), then
according to Lemma \# 2 we contradict the assumption that
$J(x) \geq 0$. Thus, for nonnegative $J$'s, its zeroes must
coincide with the turning points. $\ \diamond$

\bigskip\noindent {\bf {Lemma \# 6: $\epsilon P < E_R S$
within the local maxima regions of $S$, and
 $\epsilon P > E_R S$ within the local minima regions of
$S$. They intersect at $S$'s inflection points.}}

This immediately follows from Eq.(4), $\epsilon P = E_R S + {\epsilon\over 2} S''$.
 We see that $\epsilon P(x)$ and $E_R S(x)$ intersect
at the inflection points, $S''(x_i) = 0$. At infinity, since
$\lim_{x\rightarrow \pm \infty} S''(x) = 0^+$, then
$\lim_{x\rightarrow \pm \infty}(\epsilon P(x) - E_R S(x)) = 0^+$.
Between any two successive extremas of $E_R S(x)$ there must
be an intersection by $\epsilon P(x).\ \diamond$

\bigskip\noindent {\bf {Lemma \# 7: Let $(V_I-E_I)$ be asymptotically
monotonically increasing, with one real turning point ($\tau$).
Let $(x_1,x_2)$ be an interval whose endpoints correspond to successive
extrema for $S$, where $S'(x_1) = S'(x_2) = 0$. 
If $(x_1,x_2) \subset (\tau,\infty)$,
and $S(x)$ is increasing within the interval, then
$P(x)$ must
be monotonically decreasing within the interval, and vice versa 
(i.e. $S \leftrightarrow P$). When $(x_1,x_2) \subset (-\infty,\tau)$, then
if $S$ is decreasing within the interval, $P$ must be
monotonically increasing within the same interval, and vice versa.}}

Under the conditions of the Lemma, $J$ is positive (by Lemma \# 3); therefore,
at any extremal value for $S$ (i.e. $S'(x_o) = 0$) , 
from Eq.(5), it follows that $\epsilon P'(x_o) = -2(V_I(x_o)-E_I) J(x_o)$.
If $x_o > \tau$, then $(V_I(x_o)-E_I) > 0$, and we have that $P'(x_o) < 0$.
Likewise, for $S'$. That is if ${\tilde x}_o > \tau$, and $P'({\tilde x}_o) = 0$,
then from Eq.(5) $E_R S'({\tilde x}_o) = -2(V_I({\tilde x}_o)-E_I)
 J({\tilde x}_o) < 0$.

Let $\tau \leq x_1 < x_2$, and let $(x_1,x_2)$ denote an open interval defined by
 two successive extremum points for $S$, within
which $S$ is increasing. It then follows that $P'(x)$ must be negative at each endpoint.
However,  $P$ cannot have a local extremum within  the interval,$P'(x_e) = 0$,
since then $S'(x_e) < 0$, contradicting the assumptions made. Thus, $P$
must be stricly decreasing within the closed interval $[x_1,x_2]$.

If $x_1 < x_2 \leq \tau$, then if $(x_1,x_2)$ denotes an open interval
on which $S$ is
decreasing ($x_{1,2}$ being two successive extremum points), then $P'(x)$ must be positive at each endpoint. $P'$ must remain positive within the interval,
otherwise at any internal extremum point, $S'(x_e) > 0$, contradicting the 
assumptions made. $\ \diamond$

>From Eq.(5), under the conditions of Lemma \# 3, then 
$\epsilon P(x) + E_R S(x)$ has only one extremum point, a
 global maximum, at the turning 
point. 

\bigskip\noindent {\bf {Lemma \# 8: If $P(x_o) = 0$, then 
$S(x)$ has a local maximum at $x_o$ }}

If $P(x_o) = 0$, then $P'(x_o) = 0$ and $J(x_o) = 0$. From Eq.(5),
$S'(x_o) = 0$. From Eq.(4) we have that $\epsilon S''(x_o) + 2E_R S(x_o) = 0$.
This cannot be satisfied at any local minimum (since there,
one has $S'' > 0$, and $S \geq 0$). We cannot have $S'' = 0$
and $S = 0$, since both $S$ and $P$ cannot be simultaneously 
0. The only possibility is that $x_o$ is a local maximum 
for $S \ \diamond$.

\bigskip\noindent {\bf {Lemma \# 9: If $S(x_o) = 0$, then
$P(x)$ has a local maximum at $x_o$}}

If $S(x_o) = 0$, then $S'(x_o) = 0$ and $J(x_o) = 0$. From Eq.(5), 
$P'(x_o) = 0$.
>From Eq.(6) $J'(x_o) = 0$. If we differentiate Eq.(5), 
$\epsilon P''(x_o) = -E_R S''(x_o)$. From Eq.(4), $2 P(x_o) = S''(x_o) > 0$,
therefore $P''(x_o) < 0$, corresponding to $P$ having a local
maximum at $x_o. \ \diamond$.

The preceding Lemmas do not shed any immediate resolution to the
question as to when is $E_I = 0$. This is because they are mostly 
of a local nature and do not make  use of the boundedness 
criteria, for physical solutions, other than Lemma's \#3 and \#4.
 Nevertheless, these represent two important
contributions. 

A potentially significant question is, for what $(\alpha,E_R,E_I)$
values will the quantization of $J \geq 0$, through EMM, 
yield any results? The impact of $J$'s nonnegativity on the reality 
of $E$ is presently under investigation.
  
For completeness, we
note that if $E_I = 0$, and $V_I$ is asymptotically monotonic, with
one zero point, then $J > 0$, as is evident from Eq.(6).

\subsection{Additional Properties}

Let $V(x)$ be an arbitrary, ${\cal P}{\cal T}$ invariant, potential: $V^*(-x) = V(x)$. For any  bounded, complex, solution
 to the Schrodinger equation, $\Psi(x)$,
with energy, $E$,  the configuration $\Psi^*(-x)$ solves the same quantum problem but with eigenenergy $E^*$. This means that both $\{S(x),P(x),J(x)\}$ and
$\{S(-x),P(-x),J(-x)\}$ solve Eqs.(10-12) for $(E_R,E_I)$ and $(E_R,-E_I)$,
respectively. That is $\{u_p,v_p,w_p\} \leftrightarrow \{(-1)^pu_p,(-1)^pu_p,(-1)^pu_p\}$, satisfy the coupled moment equations (together with $(E_R,E_I) \leftrightarrow (E_R,-E_I)$).

There is another perspective on the above. The $\Psi^*(x)$ solves the equation
\begin{equation}
 \epsilon\partial_x^2\Psi^*(x) + V(x) \Psi^*(x) = -E^*\Psi^*(x).
\end{equation}
Thus, we may regard $\Psi^*(x)$ as the solution to the Schrodinger equation upon taking $\epsilon \rightarrow -\epsilon$ (with eigenenergy $-E^*$). Similarly,
$S(x), P(x), -J(x)$ solves Eqs.(10-12) for $\epsilon \rightarrow -\epsilon$,
 $E_R \rightarrow -E_R$, and $E_I \rightarrow E_I$.

The Eigenvalue Moment Method is an $L^1$ quantization theory which is analytic in $\epsilon$. That is, bounded configurations are normalized according to
$L^1$ integral relations (i.e. $\int dx \ S(x) = 1$, etc.) and not the usual 
$L^2$ conditions within the usual quantum mechanics Hilbert space formulation.
In the present case, since we are working with $\{S,P,J\}$, our $L^1$ 
normalization coincides with the $L^2$ formulation of quantum mechanics, so 
long as $\epsilon \neq 0$. Within the usual quantum mechanical formalism, the
eigenstates of the position operator (i.e. the translated Dirac delta function)
are non-normalizable (within the $L^2$ norm). However, they are normalizable
within the $L^1$ norm inherent to the EMM approach.

The final observation is that upon performing the change of variables $y = {x\over s}$,
the Hamiltonian under consideration becomes
\begin{equation}
-{\epsilon\over {s^2}}\partial_y^2\Psi + i(s^3y^3 + \alpha s y)\Psi = E \Psi.
\end{equation}
Dividing by $s^3$, we see that
\begin{equation}
E({\alpha\over {s^2}},{\epsilon\over{s^5}}) = {{E(\alpha,\epsilon)}\over {s^3}}.
\end{equation}
So it follows that if $ s = \epsilon^{1\over 5}$, then

\begin{equation} E(\alpha,\epsilon) = 
\epsilon^{3\over 5} E({\alpha\over {\epsilon^{2\over 5}}},1).
\end{equation}
\vfil\break
\section{ EMM-Numerical Analysis of the $P^2+iX^3+\alpha i X$ Hamiltonian}

The recursive, linear structure, of the $u$-moment equation in Eq.(13) can
be written as (i.e. $\epsilon = 1$)

\begin{equation}
u_p = \sum_{\ell = 0}^6 {\tilde M}_{p,\ell}(E_R,E_I) u_\ell,
\end{equation}
for $p \geq 0$, where ${\tilde M}_{p,\ell}$ satisfies Eq.(13) with respect to the $p$ index, as well as the initialization conditions ${\tilde M}_{\ell_1,\ell_2} = \delta_{\ell_1,\ell_2}$, for $0 \leq \ell_{1,2} \leq 6$.

These relations are supplemented by Eq.(14), written in the form
$u_3 = E_I u_0 - \alpha u_1$.
We can then substitute in the previous relation, obtaining
\begin{equation}
u_p = \sum_{\ell = 0 }^6 M_{p,\ell} u_\ell,
\end{equation}
where 
\begin{equation}
M_{p,\ell } = \cases { 
	                     {\tilde M}_{p,0}+ E_I {\tilde M}_{p,3}, \  \ \ell = 0 \cr
			 {\tilde M}_{p,1}-\alpha {\tilde M}_{p,3}, \  \ \ell = 1 \cr 
			{\tilde M}_{p,2}, \  \ \ell = 2 \cr
				0, \  \ \ell = 3 \cr
		        {\tilde M}_{p,\ell}, \  \ 4 \leq \ell \leq 6	}
\end{equation}

Since $M_{p,3} = 0$, we will work with the reduced set of independent moment variables $\nu_0 = u_0$, $\nu_1 = u_1$, $\nu_2 = u_2$, $\nu_3 = u_4$,
$\nu_4 = u_5$, and $\nu_5 = u_6$. Thus

\begin{equation}
u_p = \sum_{\ell = 0}^5 \Omega_{p,\ell} \nu_\ell,
\end{equation}
where $\Omega_{p,\ell} = M_{p,\ell}$, for $0 \leq \ell \leq 2$, and
$\Omega_{p,\ell} = M_{p,\ell+1}$, for $3 \leq \ell \leq 5$.

Finally, we define our normalization condition with respect to the even order $u_p$ moments $\sum_{\ell = 0}^5 u_{2\ell} = 1$. Since each of these is positive,
it insures that the requisite linear programming analysis  is done within 
the five dimensional unit-hypercube $[0,1]^5$.

 In order to impose this normalization condition,
 we invert the $\nu_\ell \leftrightarrow u_{2\ell}$ relation
(i.e. $u_{2\ell} = \sum_{\ell_v = 0}^5 \Omega_{2\ell,\ell_v} \nu_{\ell_v}$,
$0 \leq \ell \leq 5$)

\begin{equation}
\nu_\ell = \sum_{\ell_v=0}^5 N_{\ell,\ell_v} u_{2 \ell_v},
\end{equation}
$0 \leq \ell \leq 5$, and substitute into Eq.(34),
$u_p = \sum_{\ell = 0}^5 \Omega_{p,\ell} \Big( \sum_{\ell_v=0}^5 N_{\ell,\ell_v} u_{2 \ell_v}\Big )$ obtaining

\begin{equation}
u_p = \sum_{\ell = 0}^5 \Gamma_{p,\ell} u_{2\ell},
\end{equation}
where $\Gamma_{p,\ell} = \sum_{\ell_v = 0}^5 \Omega_{p,\ell_v} N_{\ell_v,\ell}$.
We now insert the normalization condition (i.e. solve for $u_0$ in terms of
the first five even order moments), obtaining
\begin{equation}
u_p = \Gamma_{p,0} + \sum_{\ell = 1}^5 \Big( \Gamma_{p,\ell}-\Gamma_{p,0}\Big) u_{2\ell},
\end{equation}
for $p \geq 0$. The linear programming EMM algorithm is implemented on these
relations, within the context of Eq.(24), or the equivalent quadratic form
counterpart to Eq.(25).

The numerical results of our analysis are given in Table I, for a selected number of $\alpha$ values, of interest within the asymptotic analysis by 
Delabaere and Trinh (2000). The reasons we do not quote more bounds,
for more $\alpha$ values, is that the Multiscale Reference Function (MRF)
analysis of Handy, Khan, Wang, and Tymczak (HKWT, 2001) is numerically faster, and
yields results lying within the EMM bounds given here, for the selected 
$\alpha$ values. This strongly suggests that the MRF analysis is correct.
This is a good example of how the present ``bounding" method can be used to test other (generally much faster)
estimation methods.

The MRF approach predicts that, for the lowest lying discrete states, there
is a critical $\alpha$ value
 below which complex energies appear. This is given by

\begin{equation}
\alpha_{critical} = -2.6118094.
\end{equation}

The data given in the Tables (particularly Table II)
 is meant to test the reliability of this.
It is clear that it does.

The  bounds for the real and imaginary parts of
the eigenenergies are very good. In the Tables, $P_{max}$ defines the maximum
moment order used. That is, it is
 the total number of Hamburger moments used (i.e.
$\{\mu_p|0 \leq p \leq P_{max}\}$). 
 If the EMM procedure is applied to the $E_I = 0$ case (i.e. Handy (2001)), 
corresponding to symmetric $S(x)$'s, then the formalism converts to a Stieltjes 
moment representation which only involves the even order moments:
$\{\mu_{2\rho}| 0 \leq \rho \leq P_{max}^{(S)}\}$. Thus, a Hamburger 
moment order of $P_{max}$, corresponds to 
a Stieltjes moment order of $ P_{max}^{(S)} ={{ P_{max}}\over 2}$. 
The tight bounds in Handy's original work (2001) required $P_{max}^{(S)} = 
O(60)$.

\begin{table}
\caption {Bounds for the Discrete States of $P^2 + iX^3 + i\alpha X$}
\begin{center}
\begin{tabular}{cccccl}
\multicolumn{1}{c}{$\alpha$}& \multicolumn{1}{c}{$P_{max}$} & \multicolumn{1}{c}{$E_R^{(L)} < E_R  < E_R^{(U)}$} & \multicolumn{1}{c}{$E_I^{(L)} < E_I< E_I^{(U)}$}\\ \hline
-3 & 20 & $0.7 < E_R < 1.7$  & $0.4 < \pm E_I < 1.0$ \\
-3 & 24 & $1.10< E_R < 1.45$ & $ 0.5 < \pm E_I < 0.9$ \\
-3 & 28 & $1.20< E_R < 1.23$ & $ 0.72 < \pm E_I < 0.77$ \\
-3 & 32 & $1.219 < E_R < 1.230$ &   $0.756 < \pm E_I <0.768$ \\
-3 & 36 & $1.224 < E_R < 1.228$ &   $0.758 < \pm E_I <0.762$ \\
-3 & 40 & $1.22561< E_R < 1.22608$ & $0.75980 < \pm E_I < 0.76055$ \\
-3 &    &  $    1.225844^*$ & $                .760030^*$\\
-2 & 20 & $.416 < E_R < .719 $ & $-.5 < E_I < .5 $ \\
-2 & 24 & $.607 < E_R < .636 $ & $-.03 < E_I < .03 $ \\
-2 & 28 & $.610 < E_R < .625 $ & $ -.5\times 10^{-2}< E_I < .5\times 10^{-2}$ \\
-2 & 32 & $.619 < E_R < .625 $ & $-.2\times 10^{-2}< E_I <.2 \times 10^{-2}$ \\
-2 & 36 & $.6203 < E_R < .6213$ & $-.45\times 10^{-3}< E_I <.45\times 10^{-3}$ \\
-2 & 40 & $.62083 < E_R < .62105 $ & $- 10^{-4}< E_I < 10^{-4}$ \\ 
-2 &    &       $0.6209137^*$ & $0^*$ 
\end{tabular}
\end{center}
\noindent{*Multiscale Reference Function formulation by Handy, Khan, Wang, and
Tymczak (2001)}
\end{table}

\begin{table}
\caption {Bounds for the Discrete States of $P^2 + iX^3 + i\alpha X$}
\begin{center}
\begin{tabular}{cccccl}
\multicolumn{1}{c}{$\alpha$}& \multicolumn{1}{c}{$P_{max}$} & \multicolumn{1}{c}
{$E_R^{(L)} < E_R  < E_R^{(U)}$} & \multicolumn{1}{c}{$E_I^{(L)} < E_I< E_I^{(U)
}$}\\ \hline
-2.610 & 20 & $.517 < E_R < 1.920$ & $-.6 < E_I < .6$\\
-2.610 & 24 & $ .940 < E_R < 1.800$ &  $-.4 < E_I < .4$\\
-2.610 & 28 & $1.083 < E_R < 1.586 $ & $-.2 < E_I < .2 $\\
-2.610 & 32 & $1.211 < E_R < 1.361$ & $-.08 < E_I < .08 $\\
-2.610 & 36 & $1.214 < E_R < 1.250$ & $-.23 \times 10^{-1} < E_I < .23 \times 10^{-1} $\\
-2.610 & 40 & $1.2135 < E_R < 1.2617$ , $1.2617 < E_R < 1.3581$ & $-.5\times 10^{-2} < E_I < .5\times 10^{-2}$ \\
-2.610 & 42 & $1.2317 < E_R < 1.2367$ , $1.3179 < E_R < 1.3356$ & 
$-.25\times 10^{-2} < E_I < .25\times 10^{-2}$ \\
-2.610 & 42  & 1.234216$^*$ and 1.332059$^*$    &  0$^*$\\                     
-2.614 & 20 & $.515 < E_R < 1.925$ & $-.525 < E_I < .525$ \\
-2.614 & 24 & $.953 < E_R < 1.808$ & $-.39 < E_I < .39 $\\
-2.614 & 28 & $1.083 < E_R < 1.589$ & $-.12 < E_I < .12 $ \\
-2.614 & 32 & $1.238 < E_R < 1.326$ & $ .01 < \pm E_I < .11$ \\
-2.614 & 36 & $1.256 < E_R < 1.309$ & $ .030 < \pm E_I < .065 $\\
-2.614 & 40 & $1.278 < E_R < 1.286$ & $ .050 < \pm E_I < .065 $ \\
-2.614 & 40   & 1.282333$^*$  & .0538739$^*$ 
\end{tabular}
\end{center}
\noindent{*Multiscale Reference Function formulation by Handy, Khan, Wang, and
Tymczak (2001)}
\end{table}

Our results (combined with those of HKWT (2001)) 
confirm the asymptotic estimates provided by Delabaere and Trinh.
In particular, we can access regions in the $\alpha$ parameter space which
were difficult within their formulation.

The numerical implementation given, of the present formalism, is meant to  
suggest the power of the method. Our intention is not to present an exhaustive
numerical analysis over a wide range of parameter values.

\subsection{ Basic Algebraic Structure of the $u$-Moments}

Upon using Eq.(14) to solve for $u_3$ in terms of $u_{0,1}$, and incorporating
this into the moment recursion equation in Eq.(13), we obtain

\begin{equation}
u_8 = \Big( {{25\alpha E_I^2}\over 6} + 5 \epsilon E_R\Big ) u_0
-{{25 \alpha^2 E_I}\over 6} u_1 - {{5E_I^2}\over 2} u_2 
-{{5 \alpha^2}\over 3 } u_4 + {{7E_I}\over 2} u_5 - {{8\alpha}\over 3} u_6,
\end{equation}
and
\begin{eqnarray}
u_{10} = \Big( -{{31 \alpha^2 E_I^2}\over 2} + {{21\epsilon^2}\over 2} -12\alpha \epsilon E_R\Big) u_0 + \Big({{31 \alpha^3 E_I}\over 2} - {{11 E_I^3}} \Big ) 
u_1\cr + \Big ( {{45\alpha E_I^2}\over 2} + 21 \epsilon E_R \Big) u_2 
+\Big( 4\alpha^3 + 12 E_I^2\Big) u_4  - {{27\alpha E_I}\over 2} u_5 + 5\alpha^2 u_6.
\end{eqnarray}

In obtaining Eq.(36), we must invert the relationship $\{u_8,u_{10}\} \leftrightarrow \{u_1,u_5\}$. This takes on the form

\begin{equation}
\pmatrix{u_8 - R_8[u_0,u_2,u_4,u_6] \cr u_{10}-R_{10}[u_0,u_2,u_4,u_6] \cr}
 = \pmatrix{  -{{25 \alpha^2 E_I}\over 6}  , {{7E_I}\over 2} \cr
          {{31 \alpha^3 E_I}\over 2} - {{11 E_I^3}} , - {{27\alpha E_I}\over 2} \cr} \pmatrix{u_1 \cr u_5\cr},
\end{equation}
where $R_{8,10}$ denote the remainder terms in the previous relations.
The determinant of the above matrix is

\begin{equation}
Det(E_I) \equiv || \pmatrix{  -{{25 \alpha^2 E_I}\over 6}  , {{7E_I}\over 2} \cr
          {{31 \alpha^3 E_I}\over 2} - {{11 E_I^3}} , - {{27\alpha E_I}\over 2}
\cr}|| = 2\alpha^3E_I^2 + {{77\over 2}}E_I^4.
\end{equation}
Accordingly,
\begin{equation}
\pmatrix{u_1 \cr u_5\cr} =
{1\over{Det(E_I)}} \pmatrix{ - {{27\alpha E_I}\over 2} , - {{7E_I}\over 2} \cr
	-({{31 \alpha^3 E_I}\over 2} - {{11 E_I^3}}), -{{25 \alpha^2 E_I}\over 6} \cr } \pmatrix{u_8 - R_8[u_0,u_2,u_4,u_6] \cr u_{10}-R_{10}[u_0,u_2,u_4,u_6] \cr}.
\end{equation}
Thus, when $Det(E_I) = 0$, the even order moments $\{u_0,u_2,u_4,u_6,u_8,u_{10}\}$
are constrained to satisfy two additional relations. That is, the effective
dimension of the system drops from $6 \rightarrow 5$ (before imposing a normalization constraint).  The impact of this on the EMM bounds is unclear, at the
present time. However, a convenient feature is that $Det(E_I) \geq 0$, under
the conditions of Lemma $\# 4$.

\vfil\break
\section {Conclusion}
We have presented a  nonnegativity representation
 formalism,  amenable to the
 Eigenvalue Moment Method, for generating converging bounds to the
(complex) discrete eigenenergies of one dimensional, ${\cal P}{\cal T}$-
invariant Hamiltonians. Our analyis was presented within the 
specific context of the $H_\alpha = P^2 + ix^3 + i\alpha x$ Hamiltonian,
previously analyzed by Delabaere and Trinh.
 Our formalism readily confirms (numerically) the existence 
of both symmetry breaking solutions, and symmetry invariant solutions.
The preliminary results given have focused on the properties of the
low lying states near the first symmetry breaking bifurcation point,
with respect to the $\alpha$ parameter, as predicted by the MRF 
eigenenergy estimation method of Tymczak et al (1998a,b). Our 
preliminary, yet highly accurate, numerical results are consistent
with the asymptotic analysis methods of Delabaere and Trinh.

\vfil\break
\section {Acknowledgments}
This work was supported through a grant from the National Science Foundation
(HRD 9632844) through the Center for Theoretical Studies of Physical Systems
(CTSPS). The author extends his appreciation to Professors D. Bessis,
G. Japaridze, G. A. Mezincescu, and Xiao-Qian Wang for useful discussions.

\vfil\break

\section{ Appendix: EMM Analysis of Eq.(1)}

In this section we focus on the structure of Eq.(1) and the moment equation
resulting from it. This is not the most efficient approach, as indicated
earlier. Instead, it is best to derive the moment equation for $S$ by
first working with the coupled $S,P,J$ equations, deriving the corresponding
coupled moment equations, and then reducing these to one moment
 equation for $S$.
Nevertheless, for the sake of completeness, we outline the issues that 
need to be addressed if one wants to go from Eq.(1), directly, to a moment
equation.

The fourth order equation for the $p^2 + ix^3+\alpha i x$ Hamiltonian is

\begin{equation}
{1\over {\Lambda(x)}} S^{(4)}(x) + 
{{\alpha + 3 x^2}\over {\Lambda^2(x)}} S^{(3)}(x)
+{{4 E_R}\over{\Lambda(x)}} S^{(2)}(x) + 
+ {{4 E_R(\alpha + 3x^2)}\over {\Lambda^2(x)}} S^{(1)}(x)
-4 \Lambda(x) S(x) = 0,
\end{equation}
where $\Lambda(x) = E_I - x(\alpha + x^2).$

The function coefficient ${1\over {\Lambda(x)}}$ introduces  singularities on the physical domain, $\Lambda(\tau) = 0$. However,
 all of the solutions to Eq.(44) will be regular (Handy (2001)). These roots are the effective
``turning points" of the differential equation. They satisfy

\begin{equation}
\tau^3 + \alpha \tau = E_I .
\end{equation}
The extremal points of the function $V_I(x) \equiv x^3 + \alpha x$ satisfy  
$ x_e^2 = -{\alpha\over 3}$. 

If $\alpha \geq 0$, then there is only one real root to Eq.(45). If $\alpha < 0$, then there could be one or three roots, depending on $E_I$. That is, $x_e^{(\pm)} = \pm \sqrt{{|\alpha|}\over 3}$, and
$V_I(x_e^{(\pm)}) =\mp \beta$, where $\beta \equiv {{2|\alpha|^{3\over 2}}\over{3\sqrt{3}}}$.
If $|E_I| \leq \beta$, or $|E_I| \leq  .3849 |\alpha|^{3\over 2}$, then there
are three roots.

It is best to translate Eq.(44) to any one of the real $\tau$ roots, in order
to simplify the structure of the ensuing moment equation. Therefore,
we will work with the translated variable, $ \xi = x - \tau$. Let us define the power moments

\begin{equation}
\mu_p = \int_{\cal C}d\xi\ \xi^p S(\xi),
\end{equation}
where ${\cal C}$ is a contour in the complex-$\xi$ plane that sits on top of
the real-$\xi$ axis and deviates around the origin. Since all the solutions
(physical or not) to Eq.(44) are regular in $\xi$, we see that if $p \geq 0$,
then for the physical solutions, we can deform the ${\cal C}$ contour to
be identical to the real-$\xi$ axis. If $p < 0$, we cannot do this, and
must retain ${\cal C}$. The $p \geq 0$ power moments are referred to as
the Hamburger moments.

We note that $\Lambda(\xi) = \xi \Upsilon(\xi)$, where 
$\Upsilon(\xi) = 3\tau^2-\alpha + 3\tau \xi + \xi^2$. If there are three, real,
$\tau$-roots, then $\Upsilon(\xi)$ has two zeroes along the real axis. If there is only one
$\tau$ root, then $\Upsilon$ has no zeroes along the real axis.

Multiplying Eq.(44) (translated by an amount $\tau$) by 
$\xi^p (\Lambda(\xi))^2$, 
and integrating by parts over the contour ${\cal C}$, we obtain a moment
equation valid for $-\infty < p < +\infty$:
\begin{eqnarray}
 -(3\tau^2+\alpha)p(4-4p -p^2 + p^3)
\mu_{p-3}
-3\tau p (-4 - p + 4p^2 + p^3)\mu_{p-2} \cr
- p(p+2)(6+12 \tau^2 E_R + 4\alpha E_R + 7 p + p^2) \mu_{p-1}
-12 \tau E_R (4+ 5 p + p^2) 
\mu_p 
-  4 E_R (12 + 8p + p^2)\mu_{p+1}\cr
+ 0 \mu_{p+2}+ 4(3\tau^2+ \alpha)^3 \mu_{p+3}
+36 \tau (3 \tau^2 + \alpha)^2 \mu_{p+4} \cr
+ 12(36 \tau^4 + 15 \alpha \tau^2 + \alpha^2) \mu_{p+5}
+ 36 (9 \tau^3 + 2 \tau \alpha) \mu_{p+6}
+(144\tau^2 + 12 \alpha) \mu_{p+7}
+ 36 \tau \mu_{p+8}
+ 4 \mu_{p+9} = 0.
\end{eqnarray}

As argued by Handy and Wang (2001), as well as Handy, Trallero, and 
Rodriguez (2001), in order for EMM to yield converging bounds,
it is important that the proper (Hamburger) moment equation uniquely correspond
to the desired system. Let us represent Eq.(44) as ${\cal O}_\xi S(\xi) = 0$.
Now consider a more general problem corresponding to ${\cal O}_\xi S(\xi) = {\cal D}(\xi)$, where the inhomogeneous term corresponds to a distribution like 
expression, supported at the zeroes of $\Lambda(\xi)$. When we multiply this
system by $\Lambda(\xi)^2$, the corresponding inhomogenous term can (effectively) disappear. If one is not careful, and solely restricts the moment index to 
nonnegative values, then the resulting moment equation cannot distinguish between the desired problem (corresponding to ${\cal D} = 0$) and those for which
${\cal D} \neq 0$. In such cases, the EMM algorithm will not generate any bounds.

In order to insure that our moment equation refers to ${\cal D} = 0$, we must
work with the moment equation evaluated for $p \geq -2$, which represents the
most singular function coefficient remaining after Eq.(44) is multiplied by 
$\Upsilon(\xi)$. In the work by Handy and Wang (2001), the moment equation for
$p = -2, -1$ yielded additional contraints for the Hamburger moments. The
same is true in the present case. Thus, the moment equation for $p = -2$
becomes
\begin{eqnarray}
 4(3\tau^2+ \alpha)^3 \mu_{1}
+36 \tau (3 \tau^2 + \alpha)^2 \mu_{2} 
+ 12(36 \tau^4 + 15 \alpha \tau^2 + \alpha^2) \mu_{3}\cr
+ 36 (9 \tau^3 + 2 \tau \alpha) \mu_{4} 
+(144\tau^2 + 12 \alpha) \mu_{5}
+ 36 \tau \mu_{6}
+ 4 \mu_{7} = \Sigma(\mu_{-2},\mu_{-4}),
\end{eqnarray}
where 
\begin{equation}
\Sigma(\mu_{-2},\mu_{-4}) = -(24 \tau E_R\mu_{-2} + 36\tau \mu_{-4}).
\end{equation}
The moment equation for $p = -1$ becomes
\begin{eqnarray}
-  20 E_R \mu_{0}
+ 4(3\tau^2+ \alpha)^3 \mu_{2}
+36 \tau (3 \tau^2 + \alpha)^2 \mu_{3} 
+ 12(36 \tau^4 + 15 \alpha \tau^2 + \alpha^2) \mu_{4} \cr
+ 36 (9 \tau^3 + 2 \tau \alpha) \mu_{5} 
+(144\tau^2 + 12 \alpha) \mu_{6}
+ 36 \tau \mu_{7}
+ 4 \mu_{8} = {{(\alpha + 3\tau^2)}\over {6\tau}}\Sigma(\mu_{-2},\mu_{-4}) .
\end{eqnarray}

By combining the previous two relations, we can express $\mu_8$ in 
terms of the Hamburger moment $\{\mu_0,\mu_1,\ldots,\mu_7\}$:

\begin{eqnarray}
4 \mu_8 + ( 34 \tau - {{2\alpha}\over{3\tau}}) \mu_7 + (6 \alpha + 126 \tau^2) \mu_6
+(252 \tau^3 + 42 \alpha \tau - {{2\alpha^2}\over \tau}) \mu_5 \cr
+ 90 \tau^2 V_I'(\tau) \mu_4 
-2(\alpha - 6\tau^2) {{(V_I'(\tau))^2}\over \tau} \mu_3 \cr - 2(V_I'(\tau))^3 \mu_2 
-{2\over{3\tau}} (V_I'(\tau))^4 \mu_1 - 20 E_R \mu_0 = 0.
\end{eqnarray}

 This is an
important additional constraint on the Hamburger moments, as noted in the
work by Handy and Wang (2001). Since the only bounded 
and positive solutions to Eq.(44) are the physical solutions, EMM will work
directly on the moment equation in Eq.(47) supplemented by the above relation
(which is essential, otherwise no bounds will be generated).

The EMM numerical implementation of the above would involve (for
fixed $\alpha$) using
$\tau$ as the variable parameter (and then computing $E_I$ from Eq.(45)),
in addition to $E_R$. Of course this means that
for some $\tau$ values, one would be recovering the same $E_I$
(i.e. those satisfying the condition $|E_I| < \beta$, defined previously);
 however, since the EMM procedure
is invariant under affine maps, this is just a redundancy. In this manner,
bounds on $\tau$ (or equivalently, $E_I$) and $E_R$ would be generated.
Our actual numerical results confirm this, and are consistent with 
the bounds quoted in the Tables.

We can enhance the above by including an additional constraint corresponding
to Eq.(14). This is obtained as follows.
>From the Schrodinger equation, we know that $E = {{\int |\Psi'(x)|^2 + i\int (x^3+\alpha x)S(x)}\over {\int S(x)}}$, or (with respect to the imaginary part 
of the energy)
\begin{equation}
E_I = {{\int ((\xi+\tau)^3+\alpha(\xi+\tau))S(\xi)}\over {\int S(\xi)}},
\end{equation}
which reduces to (i.e. from Eq.(45))
\begin{equation}
\mu_{3} + 3\tau\mu_2 + (3\tau^2+\alpha) \mu_1 = 0.
\end{equation}

\vfil\break
\section {References}
\noindent Bender C M and Boettcher S 1998 Phys. Rev. Lett. {\bf 80} 5243

\noindent Bender C M, Boettcher S, and Meisinger P N 1999, J. Math. Phys. {\bf 40} 2201

\noindent Bender C M, Boettcher S, Jones H F and Savage V M 1999 J. Phys. A:
Math. Gen. {\bf 32} 1

\noindent Bender C M, Boettcher S and Savage V M 2000 J. Math. Phys. {\bf 41} 6381

\noindent Bender C M and Wang Q 2001 J. Phys. A: Math. Gen.  34 3325

\noindent Bender C M and Orszag S A,   {\it Advanced Mathematical
Methods for
Scientists and Engineers} (New York: McGraw Hill 1978).

\noindent Caliceti E 2000 J. Phys. A: Math. Gen. {\bf 33} 3753

\noindent Chvatal V 1983 {\it Linear Programming} (Freeman, New York).

\noindent Delabaere E and Pham F 1998 Phys. Lett. {\bf A250} 25

\noindent Delabaere E and Trinh D T 2000 J. Phys. A: Math. Gen. {\bf 33} 8771

\noindent Dorey P, Dunning C, and Tateo R 2001 Preprint hep-th/0103051

\noindent Handy C R 1981 Phys. Rev. D {\bf 24}, 378

\noindent Handy C R 1987a Phys. Rev. A {\bf 36}, 4411 

\noindent Handy C R 1987b Phys. Lett. A {\bf 124}, 308 

\noindent Handy C R 2001 CAU preprint (Submitted to J. Phys. A)

\noindent Handy C R and  Bessis D 1985 Phys. Rev. Lett. {\bf 55}, 931 

\noindent Handy C R, Bessis D, and Morley T D 1988a Phys. Rev. A {\bf 37}, 4557 

\noindent Handy C R, Bessis D, Sigismondi G, and Morley T D 1988b Phys. Rev. Lett
{\bf 60}, 253 

\noindent Handy C R, Khan D,  Wang Xiao-Qian, and Tymczak C J 2001 CAU preprint (submitted to J. Phys. A)

\noindent Handy C R, Luo L, Mantica G, and Msezane A 1988c Phys. Rev. A {\bf 38}, 490

\noindent Handy C R and Wang Xiao-Qian Z 2001, CAU preprint

\noindent Handy C R, Trallero-Giner C and Rodriguez A Z 2001 CAU preprint (submitted to J. Phys. A)

\noindent Levai G and Znojil M 2000 J. Phys. A: Math. Gen. {\bf 33} 7165

\noindent Mezincescu G A 2000 J. Phys. A: Math. Gen. {\bf 33} 4911

\noindent Mezincescu G A 2001 J. Phys. A: Math. Gen. {\bf 34} 3329

\noindent Shohat J A and  Tamarkin J D, {\it The Problem of Moments}
(American Mathematical Society, Providence, RI, 1963).

\noindent Shin K C 2000 Preprint math-ph/0007006 (J. Math. Phys, under press)

\noindent Tymczak C J, Japaridze G S, Handy C R, and Wang Xiao-Qian 1998 Phys. Rev. Lett. 80, 3678 

\noindent Tymczak C J, Japaridze G S, Handy C R, and Wang Xiao-Qian 1998 Phys. Rev. A58, 2708 

\noindent Znojil M 2000 J. Phys. A: Math. Gen. {\bf 33} 6825

\end{document}